# Surrogate modelling and uncertainty quantification based on multi-fidelity deep neural network

## Zhihui Li, Francesco Montomoli

UQ Laboratory, Department of Aeronautics, Faculty of Engineering, Imperial College London, London, SW7 2AZ, UK
Email address: zhihui.li@imperial.ac.uk (Zhihui Li), f.montomoli@imperial.ac.uk (Francesco Montomoli)

**Abstract**

Deep neural network (DNN) generally needs a high number of high-fidelity (HF) training data to ensure its prediction accuracy. However, in practical engineering applications, the availability of HF data is computationally or experimentally limited, mainly for the associated costs. To reduce training costs, several DNNs that can learn from a small set of HF data and a sufficient number of low-fidelity (LF) data have been proposed. In these established neural networks, a parallel structure is commonly proposed to separately approximate the non-linear and linear correlation between the HF- and LF data. In this paper, a new architecture of multi-fidelity deep neural network (MF-DNN) was proposed where one sub-network was built to approximate both the non-linear and linear correlation simultaneously. Rather than manually allocating the output weights for the paralleled linear and nonlinear correction networks, the proposed MF-DNN can autonomously learn arbitrary correlation. The prediction accuracy of the proposed MF-DNN was firstly demonstrated by approximating the 1-, 32- and 100-dimensional benchmark functions with either the linear or non-linear correlation. The surrogating modelling results revealed that MF-DNN exhibited excellent approximation capabilities for the test functions. Subsequently, the MF-DNN was deployed to simulate the 1-, 32- and 100-dimensional aleatory uncertainty propagation progress with the influence of either the uniform or Gaussian distributions of input uncertainties. The uncertainty quantification (UQ) results validated that the MF-DNN efficiently predicted the probability density distributions of quantities of interest (QoI) as well as the statistical moments without significant compromise of accuracy. MF-DNN was also deployed to model the physical flow of turbine vane LS89. The distributions of isentropic Mach number were well-predicted by MF-DNN based on the 2D Euler flow field and few experimental measurement data points. The proposed MF-DNN should be promising in solving UQ and robust optimization problems in practical engineering applications with multi-fidelity data sources.

## 1. Introduction

Deep learning technique has been applied recently to a wide range of applications to model physical [1-2], communications [3] and biological systems [4]. Generally, deep learning algorithms show an obvious data-hungry character in the data-training process, which limits its further practical applications. For example, the traditional fully connected deep neural network (DNN) needs much high-fidelity (HF) training data to achieve satisfactory prediction accuracy. However, in engineering applications, generating enough HF results often means computationally or experimentally prohibitive costs. The actual situation is that the HF data is inadequate in training the networks and the prediction accuracy is degraded accordingly. This problem also becomes crucial in the field of high-dimensional uncertainty quantification (UQ) where the input uncertainty effects on the reference system outputs need to be described and quantified. For instance, to solve the 32-dimensional UQ problem using traditional polynomial chaos expansion method, with a polynomial order of 5, around 435,897 HF samples are needed, which is computationally expensive and unaffordable for engineers.

To reduce such dependence on HF results, multi-fidelity (MF) approaches have attracted much more attention recently in surrogate modelling and UQ, leveraging a set of multi-fidelity HF and LF data sets [5-8]. Given that LF data is abundant and cheap to obtain, the fundamental idea of MF methodology is to ensure the model's prediction accuracy by increasing the amount of LF data [9]. As MF surrogate modelling, different approaches have been proposed in literature, including MF response surface models [10-11], MF artificial neural networks [12-13] and MF Gaussian process [14-18]. As for the MF UQ approaches, MF polynomial chaos expansion [19-20] and MF Monte Carlo method [21-22] were also proposed in previous research works. However, the abovementioned approaches generally show weak performance in constructing the cross-correlation between HF- and LF data. For example, based on our experience the selection of HF point locations influences significantly the accuracy of the prediction of the MF response surface model. Besides that, MF Gaussian process regression shows some difficulties in optimizing the hyperparameters with sparse data [23] and solving the high-dimensional problems [24]. Besides, the polynomial chaos expansion method strongly depends on the regularity between the Quantities of Interest (QoI) and the input uncertainties, and its performance might be degraded in the absence of this kind of regularity, e.g., the stochastic hyperbolic problems [25]. A comprehensive review of multi-fidelity methods in uncertainty quantification, statistical inference and optimization can be found in [26].

More recently MF neural networks become promising in building surrogate models that may be used for both UQ and optimization problems. Compared to traditional function approximators, the DNN is regarded as a universal approximator at either low- or high dimensions [27]. Lu and Zhu [28] used a single fully connected neural network (NN) as a bi-fidelity surrogate model to estimate the high-fidelity proper orthogonal decomposition coefficients within the framework of the reduced-order model. Yan and Zhou [29] built an adaptive surrogate model based on the MF approach in conjunction with NN to solve the Bayesian inverse problem. The NN was used to bridge the low- and high-fidelity surrogate model in the framework of Markov chain Monte Carlo Conti et al. [30] established the MF long short-term memory networks to solve the time-dependent problems. Motamed [31] constructed two separate neural networks based on bi-fidelity training data. In detail, the first neural network was specially designed to approximate the correction function between LF- and HF data, and the output of the first neural network was then set to be the extra HF training data for the second neural network. Here the precision of these artificially added HF training data significantly determines the prediction accuracy of the whole neural network. Meng and Karniadakis [32] built the composite neural network in a more straightforward way that the LF model was trained first and the output of the LF model was then corrected based on the HF data. The proposed multi-fidelity deep neural network (MF-DNN) consists of three sub-networks, one network for LF prediction, one network for HF prediction based on the linear assumption between the LF- and HF data, and one network for HF prediction based on the non-linear assumption between them. The results showed that the composite neural network could give good accuracy in approximating several benchmark functions. The applications of the abovementioned MF-DNN (i.e., the architecture with three sub-networks built in [32]) in aerodynamic optimization can be seen in [33]. Ahn et al. [34] utilized the reduced-order model to update the low-fidelity data within the framework of composite neural networks. Guo et al. [12] conducted the MF regression with the neural networks featured by different architectures and then compared the prediction accuracy against the co-kriging model. In Guo's work, one shallow neural network layer was built to approximate the correlation between LF- and HF data. However, the results showed that its approximation capability in modelling linear correction function was somewhat degraded when compared to the parallel sub-networks structure. Besides, its performance in predicting high-dimensional problems has not been revealed.

To avoid the structure of the parallel sub-networks while simultaneously balancing the approximation capability in modelling both linear and nonlinear correction, in this paper, we proposed to build a new network architecture featured with only one correction sub-network. The following attractive features can be expected: 1) the prior assumption in allocating the weights for the linear and nonlinear sub-networks is avoided; 2) this architecture closely aligns with the underlying mathematical principles of MF methodology on which it is based; 3) the training and testing process of the MF-DNN becomes straightforward because of the simplified architecture, especially for the high-dimensional problems; 4) the new MF-DNN is much easier to be implemented in programming. As far as the authors know, this is the first attempt to solve the high-dimensional aleatory UQ problem using MF-DNN. The paper is organized as follows: the details of the new MF-DNN were detailed in Sec. 2; the test results of the surrogate modelling and UQ for several benchmark cases were then shown in Sec. 3. The main conclusions of this research work were drawn in Sec. 4. In addition, the comparison of the different activation functions in approximation capability was added in Appendix A.

## 2. Multi-fidelity Deep Neural Network

One widely used comprehensive correction [9] in bridging LF- and HF data is:

$$\hat{y}_{HF} = \rho(X) \cdot y_{LF}(X) + \delta(X) \tag{1}$$

where $\hat{y}_{HF}$ represents the model prediction values on HF data points, $\rho$ is the multiplicative correction surrogate, $y_{LF}$ represents the label values of LF data points, $\delta$ means the additive correction surrogate. Here the multiplicative correction $\rho$ could be either a constant [35] or the non-constant value [36], which represents the linear or non-linear correction between $\hat{y}_{HF}$ and $y_{LF}$. In other words, the comprehensive correction in (1) can be expressed as:

$$\hat{y}_{HF} = \mathcal{F}(y_{LF}(X), X) \tag{2}$$

where $\mathcal{F}$ can comprehensively represent both the non-linear and linear correlation between the LF- and HF data. Thus, the idea for the proposed MF-DNN is that it should consist of two sub-networks, one for the LF-DNN to approximate the values of $y_{LF}$, and one for the Correction DNN to predict the $\hat{y}_{HF}$ based on the expression (2).

The architecture of the new MF-DNN is shown in Fig. 1. Here we assume that there is a large set of LF training data $y_{LF}(X_{LF}), X_{LF} \in \mathbb{R}$ and a relatively small set of HF training data $y_{HF}(X_{HF})$, $X_{HF} \in \mathbb{R}$, $X_{HF} \subset X_{LF}$. The cost function was set to be the mean square error (MSE) between the prediction results and the actual values. The gradient information of the cost function to the network parameters could be obtained based on automatic differentiation. The prediction error of the MF-DNN was minimized by using gradient-based optimization algorithms. In detail, the ADAM and L-BFGS optimizers [37-38] were deployed here because of their better generalizing performance when compared to the other optimization methods [39-40]. To avoid overfitting, the $L_2$ regularization loss (also called Ridge regression [41]) was added to minimize the loss function by summing the squared magnitude of network weight coefficients. Here the LF-DNN and Correction DNN were trained in sequence, which is beneficial for the programming and training. The definition of the loss function for the LF-DNN is shown as follows:

$$L_l = \frac{1}{M}\sum_{i=1}^{M}\left(\widehat{y_{LF}}(X_{LF};\theta_{NN}) - y_{LF}(X_{LF})\right)^2 \tag{3}$$

$$\theta_{NN} = \{(w_h, b_h)\}_{h=1}^{H+1} \tag{4}$$

where $L_l$ represents the training loss of LF-DNN, $M$ means the number of LF training points, $\theta_{NN}$ means the set of neural network parameters, $w_h$ is the weights of $h$-th layer, $b_h$ is the biases of $h$-th layer and $H$ means the number of hidden layers. The trained LF-DNN then operated as an offline surrogate model. This implies that it remained fixed and was not retrained during the training process of the Correction DNN. Furthermore, the definition of loss function for the Correction DNN is shown as follows:

$$L_c = L_h + L_r \tag{5}$$

$$L_h = \frac{1}{P}\sum_{i=1}^{P}\left(\widehat{y_{HF}}(X_{HF};\theta_{NN}) - y_{HF}(X_{HF})\right)^2 \tag{6}$$

$$L_r = \lambda \sum w_h^2 \tag{7}$$

where $L_c$ represents the overall training loss of Correction DNN, $L_h$ is the mean square error of Correction DNN, $L_r$ represents the $L_2$ regularization loss of Correction DNN, $P$ is the number of HF training points and $\lambda$ is the control parameter of $L_2$ regularization loss. In order to enable the Correction DNN to effectively learn the disparities between the predictions generated by the LF-DNN and the corresponding HF labels, the HF variables $X_{HF}$ were also incorporated during the training phase of the Correction DNN. By including these HF variables in the training process, the Correction DNN gained the ability to comprehend and capture the differences between the LF-DNN predictions and the actual HF labels, facilitating its ability to correct and refine the predictions accordingly.

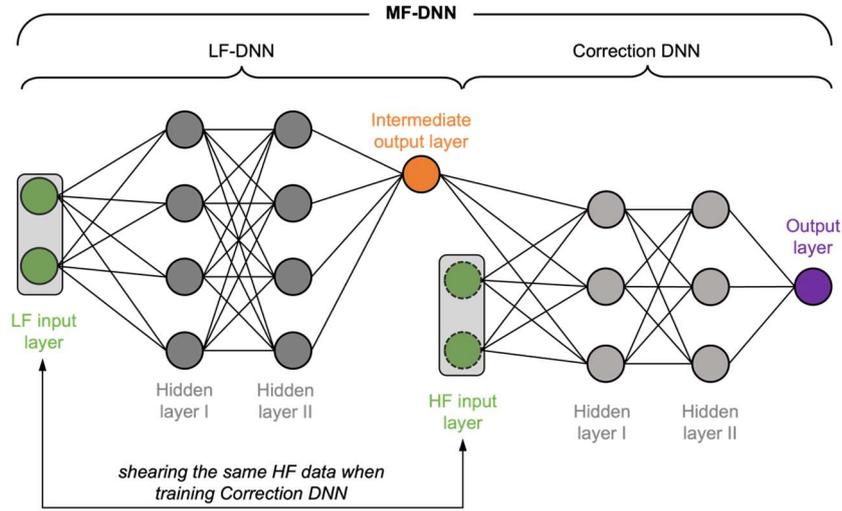

Fig. 1 Architecture of the proposed MF-DNN (taking two-layer as an example)

The network hyperparameters, i.e., the numbers of the hidden layers, the number of neurons in hidden layers, the learning rate, etc., were optimized by the Bayesian optimization algorithm [42]. The Bayesian optimization method was chosen because it is generally regarded as a global optimization method in presence of noisy black box problems, and generally shows better performance in balancing exploration and exploitation when compared to the grid search approach [43] and random search method [44]. The performance of the optimized MF-DNN was then evaluated using the K-fold cross-validation method [45]. Rather than using the commonly used hyperbolic tangent function (also known as "Tanh") in existing MF-DNNs [12, 32], the rectified linear unit activation function (also known as "ReLU") was deployed here to obtain the balance of approximating both the linear and non-linear correlation in one sub-network. The mathematical expression of "ReLU" is as follows:

$$f(x) = \max(0, x) \tag{8}$$

The "ReLU" function is beneficial for the faster and more effective training of DNNs when compared to the traditional logistic sigmoid function (also known as "Sigmoid") and the "Tanh", especially for the high-dimensional dataset [46]. More tests on the "ReLU" capability of the function approximation were shown in Appendix A at the end of this paper. The building and training of the MF-DNN were carried out using the *Keras* in the environment of *Tensorflow 2* [47]. The detailed training procedures for the MF-DNN are as follows:

1) Building the initial fully connected LF-DNN;

2) Optimizing the hyperparameters in LF-DNN using the Bayesian optimization method;

3) Training the LF-DNN based on the $M$ realizations of the LF data set $Q_{LF} = \{y_{LF}^{(1)}, \dots, y_{LF}^{(M)}\}$ using the ADAM and L-BFGS optimizers;

4) Building the initial Correction DNN and calling the trained LF-DNN in 3) as a sub-module;

5) Optimizing the hyperparameters in the Correction DNN using the Bayesian optimization method;

6) Training the Correction DNN based on the HF data set $Q_{HF} = \{y_{HF}^{(1)}, \dots, y_{HF}^{(N)}\}$ $(M \gg N)$ using the ADAM and L-BFGS optimizers;

7) Validating and testing the accuracy of the trained MF-DNN based on the K-fold cross-validation method.

## 3. Results and discussion

In this section, several benchmark tests were conducted to verify the performance of the built MF-DNN in surrogate modelling and uncertainty quantification.

*3.1 1-dimensional function with linear correlation*

Here one-dimensional function featured by the linear correlation between the LF- and HF data was tested first. The theoretical expressions of this function are as follows:

$$y_L(x) = 0.5y_H + 10(x - 0.5) - 5 \tag{9}$$

$$y_H(x) = (6x - 2)^2 \sin(12x - 4) \tag{10}$$

To approximate both the LF- and HF functions, here 21 LF points (uniform distributions in interval $X_{LF} \in [0,1]$) and 4 HF points ($X_{HF} \in [0, 0.35, 0.75, 1]$) were generated and then collected as the training data for the MF-DNN. Before the training data are transferred to the MF-DNN, the learning rate, the number of hidden layers and the number of neurons in each hidden layer were tuned using Bayesian optimization. The variation range for the hidden layer number is [1, 2, 3, 4], the variation range for the neuron number in each hidden layer is [8, 16, 24, 32, 40, 48, 56, 64], and that for the learning rate is [0.01, 0.001, 0.0001], respectively. The optimization objective is to minimize the prediction loss of the MF-DNN. The training epoch number was set to be 2000. After 20 Bayesian optimization loops, the optimal architecture of the LF-DNN was converged to be 3 hidden layers featured by 64, 64 and 40 neurons in each hidden layer, respectively. The optimal architecture of the Correction DNN was featured by 1 hidden layer with 8 neurons on it. The optimal learning rate was set to be 0.001.

The weights and biases of the MF-DNN were initially updated using ADAM optimizer for the first 1000 steps. After that, the L-BFGS optimization algorithm was employed to further minimize the training loss over the next 2000 steps. The performance of the well-trained MF-DNN is shown in Fig. 2. In detail, Fig 2 (a) shows that the MF-DNN can approximate the HF function accurately based on 4 HF data points. This also validates that the built Correction DNN with the "ReLU" activation function can well approximate the linear correlation between the LF- and HF data. Furthermore, the prediction results obtained from the proposed MF-DNN (MF_DNN_ReLU) are compared with other surrogate models, namely the Radial Basis Function (RBF), Kriging (KRG), MF-DNN with composite architecture (MF_DNN_Comp, proposed in Ref. [32]), and Co-Kriging (Co-KRG). Here the RBF and KRG models were trained based on the HF data points, and the comparison results were added in Fig. 2 (b). The MSE of these models on 20 validation data points is collected in Tab. 1. Among the models under consideration, the performance of the proposed MF-DNN outperforms that of the other surrogate models, except for the co-kriging model. This finding is not entirely surprising, as Co-Kriging was originally designed based on low-dimensional linear assumptions. This demonstrates the effectiveness of neural networks with the ReLU activation function in approximating the linear correlation between LF- and HF data.

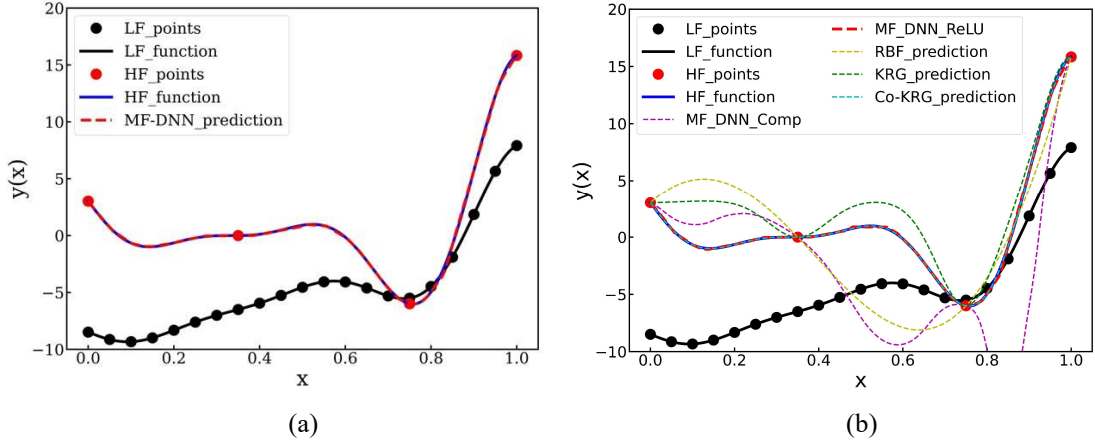

(a)                                            (b)

Fig. 2 Performance of MF-DNN in approximating 1-dimension function with linear correlation (a: Comparison between the MF-DNN prediction results and analytical values; b: Comparison of MF-DNN with the other surrogate models)

Table 1 MSE for prediction of 1-dimensional function with linear correlation

| Models | MSE of test dataset |
| --- | --- |
| MF-DNN_ReLU | 9.65e-03 |
| MF-DNN_Comp | 4.10e+00 |
| RBF | 18.39e+00 |
| KRG | 4.21e+00 |
| Co-KRG | 4.32e-09 |

### 3.2 1-dimensional function with non-linear correlation

The one-dimensional function featured by the non-linear correlation between the LF- and HF data was then tested here. The theoretical expressions of this function are as follows:

$$y_L(x) = 0.5(6x - 2)^2 \sin(12x - 4) + 10(x - 0.5) - 5 \tag{11}$$

$$y_H(x) = 0.1y_L(x)^2 + 10 \qquad (12)$$

To approximate both the LF- and HF functions, here 21 LF points (uniform distributions in interval $X_{LF} \in [0, 1]$) and 6 HF points ($X_{HF} \in [0, 0.1, 0.3, 0.7, 0.9, 1]$) were generated and then collected as the training dataset for MF-DNN. The activation function "ReLU" was deployed here to test its capability in approximating the non-linear correlation between the LF- and HF functions. Similar to the training process shown in Sec. 3.1, the hyperparameters of the neural networks were optimized by the Bayesian optimization method. The training epoch number was set to be 2000. After 20 optimization loops, the optimal architecture of the LF-DNN was featured by the 3 hidden layers with 64, 64 and 40 neurons in each hidden layer, respectively. Meanwhile, the optimal architecture of the Correction DNN was featured by 2 hidden layers with 64 and 56 neurones in each hidden layer, respectively.

The weights and biases of the MF-DNN were initially updated using the ADAM optimizer for the initial 1000 steps. Subsequently, the L-BFGS optimization algorithm was utilized to further minimize the training loss during the subsequent 2000 steps. As shown in Fig. 3 (a), the MF-DNN accurately approximates the HF function based on 6 HF data points. The performance of the MF-DNN was then compared to the RBF, KRG, MF_DNN_Comp and Co-KRG models, as shown in Fig. 3(b). In direct comparison with other models, the MF-DNN exhibits closed distributions that closely resemble the HF function. The MSE of these models in predicting 20 validation data points was collected in Tab. 2. The results indicate that the proposed MF-DNN achieves a prediction accuracy that is comparable to the composite structure used in Ref. [32]. This highlights the capability of neural networks with the ReLU activation function to effectively approximate the non-linear correlation between the LF- and HF data.

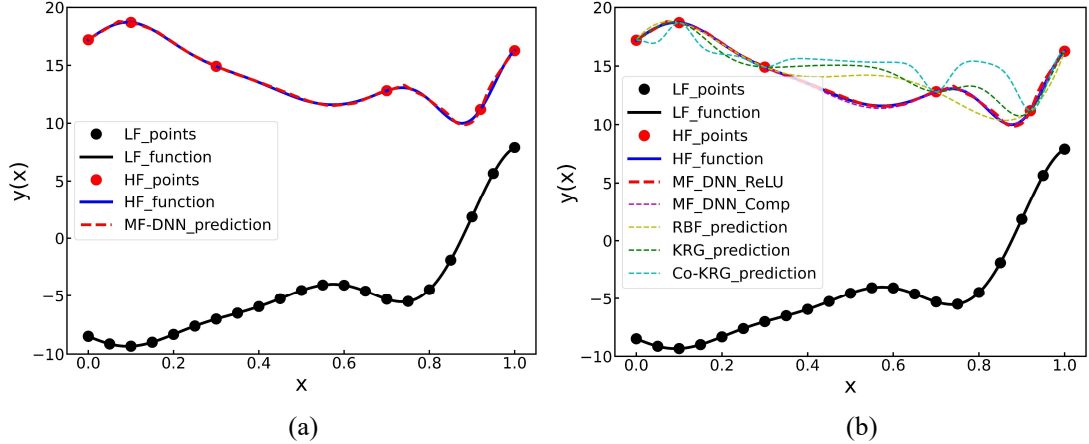

(a)                                                    (b)

Fig. 3 Performance of MF-DNNs in approximating 1-dimension function with nonlinear correlation (a: Comparison between the MF-DNN prediction results and analytical values; b: Comparison of MF-DNN with the other surrogate models)

Table 2 MSE for prediction of 1-dimensional function with non-linear correlation

| Models | MSE of test dataset |
|---|---|
| MF-DNN_ReLU | 1.15e-02 |
| MF-DNN_Comp | 1.18e-02 |
| RBF | 1.22e+00 |
| KRG | 2.05e+00 |
| Co-KRG | 4.97e+00 |

*3.3  32-dimensional function*

In many industrial design scenarios, having 32 parameters represents a relatively high-dimensional space. Thus, the capability of the proposed MF-DNN in approximating the 32-dimensional function was tested here. The theoretical expressions of this function are as follows:

$$y_L(x_0, \dots, x_{31}) = 0.8 * y_H - 0.4 \sum_{i=0}^{30}(x_i x_{i+1}) - 50, x_i \in [-3, 3] \tag{13}$$

$$y_H(x_0, \dots, x_{31}) = (x_0 - 1)^2 + \sum_{i=1}^{31}(2x_i^2 - x_{i-1})^2, x_i \in [-3, 3] \tag{14}$$

To approximate the high-dimensional function, 200,000 LF- and 2000 HF data points were generated using LHS method. The training epoch number is set to be 5000. After 20 optimization loops, the Bayesian optimization results showed that the LF-DNN consisted of 2 hidden layers with 512 neurons and 256 neurons on each layer, respectively. The Correction-DNN consisted of 1 hidden layer with 32 neurons on it. The activation function in Correction-DNN was set to ReLU, and the learning rate was set to 0.001. The training process of the MF-DNN was boosted by using the *NVIDIA TESLA K80* Graphics Processing Unit (GPU) in *Google Colaboratory*.

Initially, the weights and biases of the MF-DNN were updated using the ADAM optimizer for the first 1000 steps. Following that, the L-BFGS optimization algorithm was employed to further reduce the training loss over the next 2000 steps. Fig. 4 shows the comparison of the MF-DNN prediction results and the analytical solutions. As stated earlier, the $x$ and $y$ coordinates of the red scatter points represent the MF-DNN predictions and the corresponding analytical solutions, respectively. Ideally, if the MF-DNN prediction error is zero, the red scatter points should perfectly align with the black line. The blue points are derived from the KRG model which is only trained by the HF data. Its ability to accurately capture the underlying patterns is evident from the alignment of the red scatter points with the analytical solutions, demonstrating its effectiveness in solving high-dimensional problems. Table 3 displays the MSE of various models on a validation dataset consisting of 1000 data points. The findings demonstrate that the proposed MF-DNN model achieves a comparable level of accuracy with the composite architecture, surpassing the performance of the other evaluated models. It should be noted that the Co-KRG model fails to provide results for this problem. This limitation arises due to the considerable size of the covariance matrix ([200000, 200000]), which requires a minimum memory capacity of 320 GB to be solved.

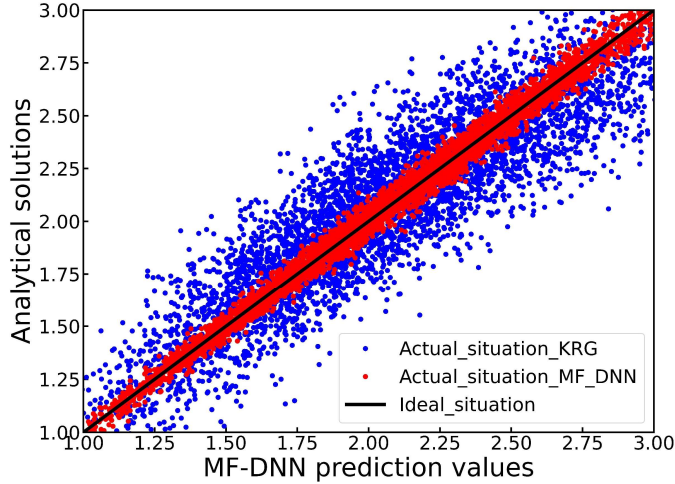

Fig. 4 Performance of MF-DNN in approximating 32-dimensional function

Table 3 MSE for prediction of 32-dimensional function with non-linear correlation

| Models | MSE of test dataset |
|---|---|
| MF-DNN_ReLU | 9.83e-04 |
| MF-DNN_Comp | 1.01e-03 |
| RBF | 1.08e+00 |
| KRG | 4.38e-02 |
| Co-KRG | NON |

*3.4 100-dimensional function*

The MF-DNN proposed in this study is evaluated by testing its performance in predicting a 100-dimensional benchmark function. The test expression of the benchmark function can be represented as follows:

$$y_L(x_0, \ldots, x_{99}) = 0.8 * y_H - 0.4 \sum_{i=0}^{98}(x_i x_{i+1}) - 50, x_i \in [-3,3] \tag{15}$$

$$y_H(x_0, \ldots, x_{99}) = (x_0 - 1)^2 + \sum_{i=1}^{99}(2x_i^2 - x_{i-1})^2, x_i \in [-3,3] \tag{16}$$

The MF-DNN model was trained using a dataset comprising 10,000,000 LF data points and 100,000 HF data points. The LF-DNN is constructed with four layers, each containing 512, 512, 256, and 128 neurons, respectively. On the other hand, the HF-DNN consists of a single layer with 64 neurons. The ReLU activation function was used in both LF-DNN and HF-DNN, and the learning rate was set to 0.001. The entire training process was carried out in the *TensorFlow 2* environment within *Google Colaboratory*.

As depicted in Figure 5, the MF-DNN demonstrates its ability to predict the analytical solutions of the 100-dimensional function effectively. The performance of various models on a validation dataset comprising 1000 scaled data points is presented in Table 4, showcasing the MSE values. The results clearly indicate that the proposed MF-DNN achieves the highest accuracy among the previous models, validating its effectiveness in addressing high-dimensional surrogate modelling problems. The RBF, KRG and Co-KRG models encounter challenges during the training process due to their significant demand for system memory, leading to an inability to complete the training successfully.

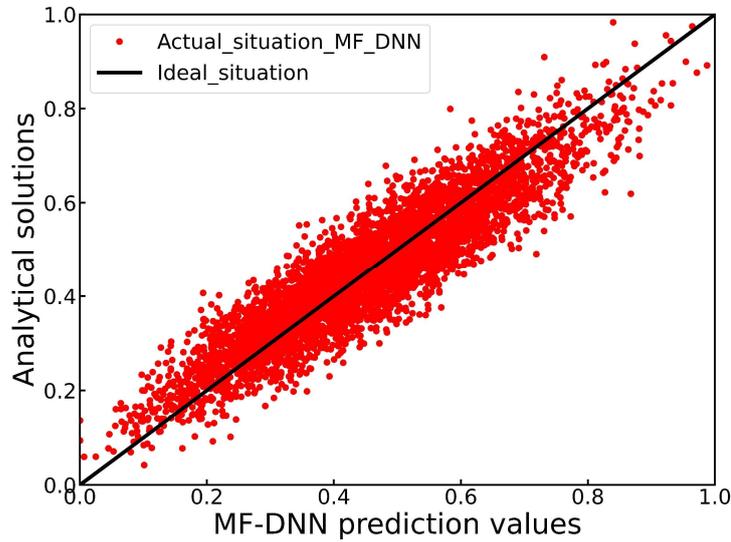

Fig. 5 Performance of MF-DNN in approximating 100-dimensional function

Table 4 MSE for prediction of 100-dimensional function with non-linear correlation

| Models | MSE of test dataset |
|---|---|
| MF-DNN_ReLU | 4.09e-03 |
| MF-DNN_Comp | 4.51e-03 |
| RBF | NON |
| KRG | NON |
| Co-KRG | NON |

## 3.5 UQ for 1-dimensional function

The influence of input aleatory uncertainties on the QoI can be evaluated based on the MF-DNN in conjunction with Monte Carlo sampling method. The expressions (9-10) were chosen as the governing equations of uncertainty propagation first, and the two representative types of input uncertainty distributions, i.e., the uniform distributions and Gaussian distributions were introduced within the UQ process. In detail, $x \in \Gamma$ is either a uniformly distributed random variable on $\Gamma = \mathrm{U}[0.6, 0.8]$ or a Gaussian distributed random variable on $\Gamma = \mathrm{N}[0.7, 0.03^2]$. The multidimensional truncated Gaussian method [49] was used to generate the Gaussian distributions within the specific bounded interval. The Monte Carlo method was selected here because of its simplicity and robustness of the implementation. The MF-DNN-based statistical moments of QoI under these two different input uncertainties were collected and compared to the analytical results, as shown in Tab. 5. MF-DNN accurately predicted the mean and variance of the QoI, and the maximal prediction error occurred in the kurtosis which was related to the tail distributions of the QoI.

Table 5 Comparison of statistical moments of QoI for 1-dimensional function

| *Statistic Description- Input uncertainties with uniform distribution* | | | | |
|---|---|---|---|---|
| | mean | variance | skewness | kurtosis |
| Analytical results | -3.926 | 3.706 | 0.549 | -1.165 |
| MF-DNNs prediction (1e+07 samples) | -3.929 | 3.723 | 0.551 | -1.165 |
| Prediction error (%) | 0.076 | 0.462 | 0.542 | 0.012 |
| *Statistic Description- Input uncertainties with Gaussian distribution* | | | | |
| | mean | variance | skewness | kurtosis |
| Analytical results | -4.426 | 1.344 | 0.691 | -0.087 |
| MF-DNNs prediction (1e+07 samples) | -4.423 | 1.356 | 0.680 | -0.090 |
| Prediction error (%) | 0.068 | 0.884 | 1.592 | 3.448 |

The histogram comparison of the QoI probability density distributions under different distributions of input uncertainties is shown in Fig. 6. The comparison results show that the QoI with the maximal probability appears near -6 for the input uncertainties with either the uniform- or the Gaussian distributions. However, the probability density of the QoI shows the aggregated distributions within the interval [-6, -4] when the input uncertainty distributions are modified from the uniform- to the Gaussian type. The overall tendency in this comparison shows that the MF-DNN can properly simulate the low-dimensional uncertainty propagation process.

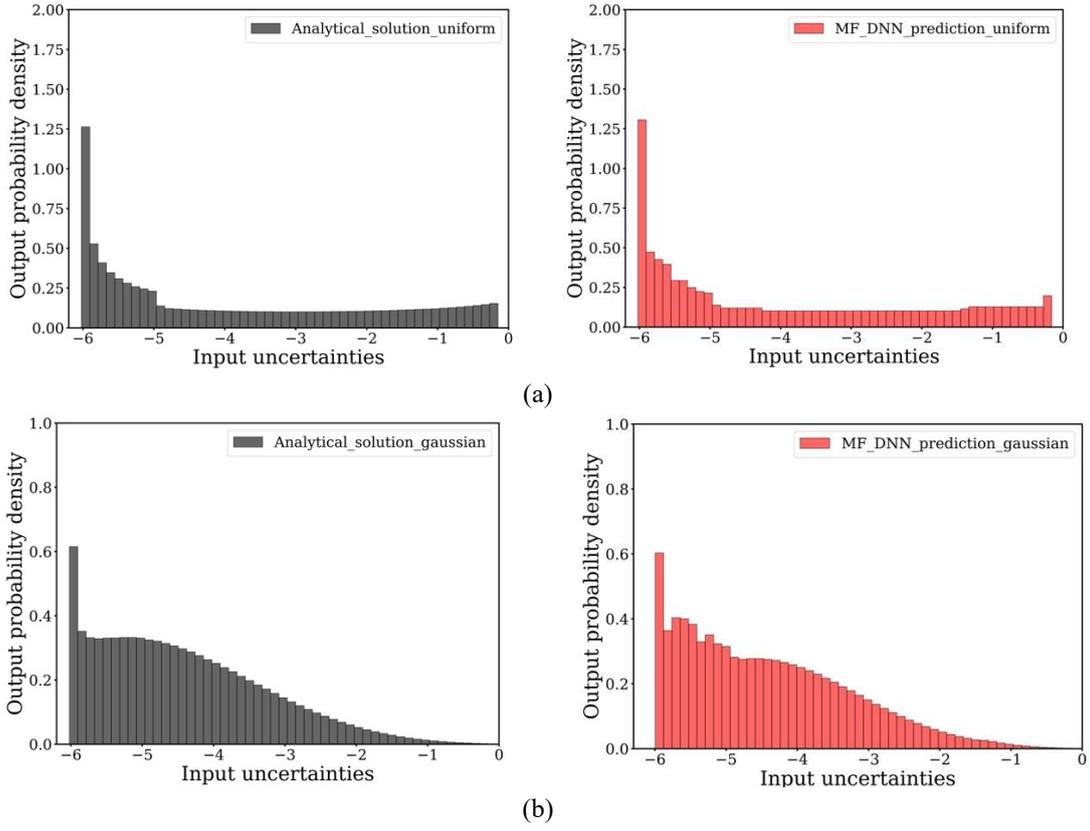

Fig. 6 Histogram comparison of QoI probability density distributions for 1-dimensional function

### 3.7 UQ for 32-dimensional function

To further test the MF-DNN capability in handling the high-dimensional UQ problem, the 32-dimensional benchmark test function (the mathematical expressions shown in equations 13-14) was chosen here as the governing equation of uncertainty propagation. The input uncertainty $x \in \Gamma$ is either a 32-dimensional uniformly distributed random variable on $\Gamma = U\left[-3, 3\right]$ or a 32-dimensional Gaussian distributed random variable on $\Gamma = N\left(0, 1^2\right)$. The multidimensional truncated Gaussian method was used to generate the normal distributions on 32 dimensions. The Monte Carlo method was deployed to simulate the uncertainty propagation process, and the MF-DNN-based statistical moments of QoI with two types of input uncertainties were collected and compared to the analytical results, as shown in Tab. 6.

Table 6 Comparison of statistical moments of QoI for 32-dimensional function

| Statistic description- Uniform distribution | | | | |
|---|---|---|---|---|
| | mean | variance | skewness | kurtosis |
| Analytical results | 2.106 | 0.269 | 0.313 | 0.103 |
| MF-DNNs predictions (1e+7 samples) | 2.105 | 0.266 | 0.300 | 0.080 |
| Prediction error (%) | 0.022 | 0.255 | 1.344 | 14.483 |

| Statistic description- Gaussian distribution | | | | |
| --- | --- | --- | --- | --- |
| | mean | variance | skewness | kurtosis |
| Analytical results | 0.364 | 0.029 | 0.931 | 1.144 |
| MF-DNNs predictions (1e+7 samples) | 0.364 | 0.029 | 0.908 | 0.939 |
| Prediction error (%) | 0.200 | 0.690 | 2.469 | 17.834 |

Although there are some errors in the kurtosis predictions, the MF-DNN shows high accuracy in predicting the mean and variance of the QoI responses. The histogram comparison of the QoI probability density distributions for the 32-dimensional function is shown in Fig. 7. The probability density distributions of the QoI are "biased" leftward when the input uncertainties vary from the uniform- to the Gaussian distribution. The overall tendency shows that the MF-DNN can qualitatively and quantitatively simulate the uncertainty propagation process for the high-dimensional UQ problem.

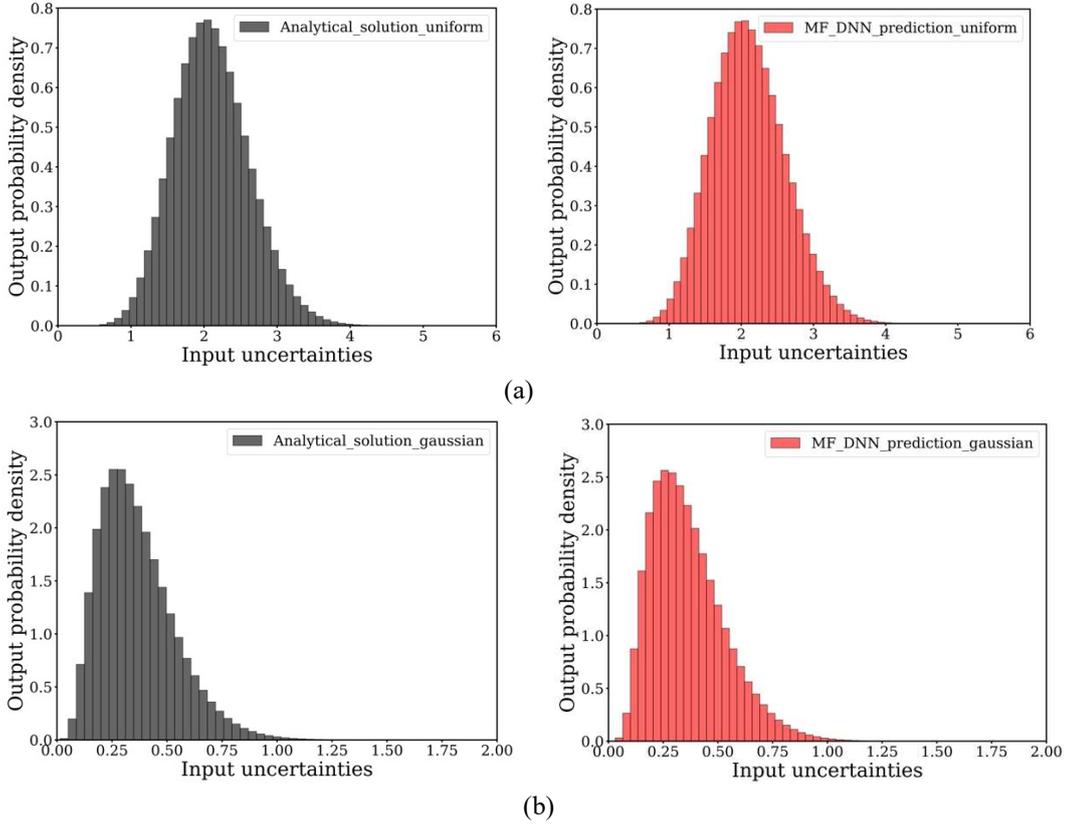

(a)

(b)

Fig. 7 Histogram comparison of QoI probability density distributions for 32-dimensional function

*3.7 UQ for 100-dimensional function*

The MF-DNN model was utilized to address the 100-dimensional UQ problem. The input uncertainty, denoted as $x \in \Gamma$ , can take two different forms: either a 100-dimensional uniformly distributed random variable on the interval $\Gamma = U[-1, 1]$, or a 100-dimensional Gaussian distributed random variable on the interval $\Gamma = N(0, 1^2)$. For both types of input uncertainties, the MF-DNN was employed to compute the statistical moments of the QoI. These computed moments were then compared to the corresponding analytical results, and the comparison outcomes are summarized in Table 7. Furthermore, the probability density distributions of the QoI were obtained using the MF-DNN, and the histogram comparison of these distributions is depicted in Fig. 8. The MF-DNN demonstrates high accuracy in predicting the mean value, while there are relatively larger discrepancies observed in the prediction of variance, skewness, and kurtosis when compared to the analytical results. Despite these discrepancies, the probability distributions derived from the MF-DNN align well with the analytical solutions, indicating a good match in capturing the overall shape and characteristics of the data distribution.

Table 7 Comparison of statistical moments of QoI for 100-dimensional function

| *Statistic description- Uniform distribution* | | | | |
| --- | --- | --- | --- | --- |
| | mean | variance | skewness | kurtosis |
| Analytical results | 6.716 | 0.861 | 0.176 | 0.029 |
| MF-DNNs predictions (1e+7 samples) | 6.723 | 0.723 | 0.159 | 0.033 |
| Prediction error (%) | 0.104 | 16.028 | 9.659 | 12.121 |
| *Statistic description- Gaussian distribution* | | | | |
| | mean | variance | skewness | kurtosis |
| Analytical results | 1.159 | 0.094 | 0.521 | 0.358 |
| MF-DNNs predictions (1e+7 samples) | 1.150 | 0.072 | 0.406 | 0.209 |
| Prediction error (%) | 0.776 | 23.400 | 22.073 | 71.291 |

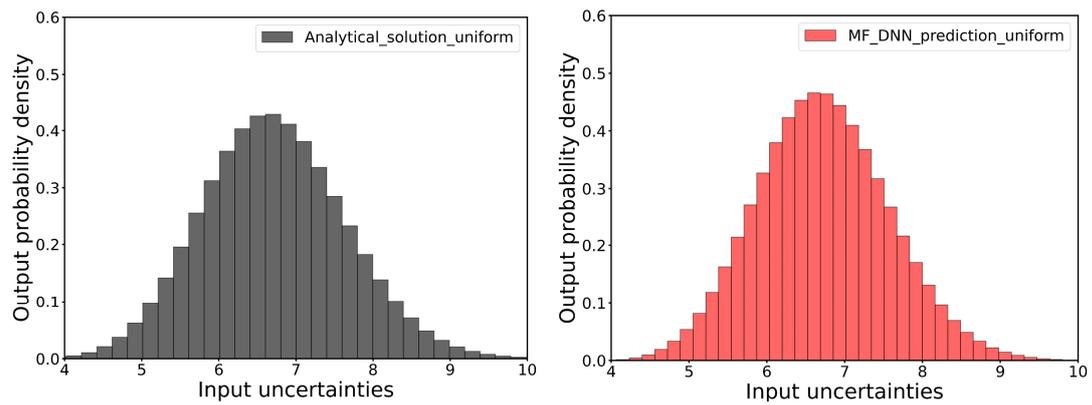

(a)

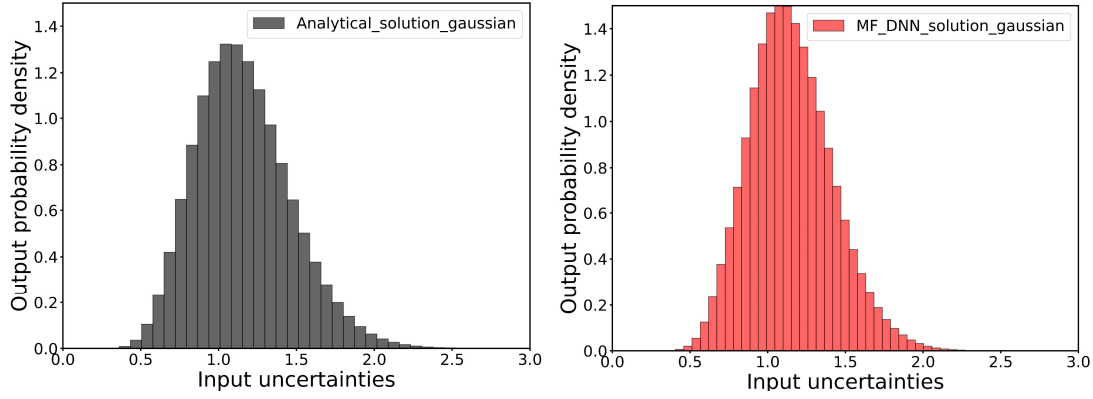

Fig. 8 Histogram comparison of QoI probability density distributions for 100-dimensional function

### 3.8 Application in predicting LS89 flow

The MF-DNN was then employed to predict the flow field of turbine value LS89. The LS89 nozzle guide vane [50] was designed and tested at the von Karman Institute for Fluid Dynamics with well-documented experiment measurements. The key geometric and operating parameters of the LS89 blade are defined in Tab. 8.

Table 8 Key parameters of LS89 vane

| Parameter | Value |
|---|---|
| Chord | 67.647 mm |
| Pitch | 57.500 mm |
| Stagger angle | 55° |
| LE radius | 4.127 mm |
| TE radius | 0.710 mm |
| Inlet total pressure | 159, 600 Pa |
| Inlet total temperature | 420 K |
| Inlet flow angle | 0° |
| Inlet turbulence intensity | 3 % |
| Outlet static pressure | 82,350 Pa |

To predict the distributions of the near-wall isentropic Mach number on the vane surface, two levels of training data were deployed here: abundant 2D Euler flow as the LF data source; few experimental measurement results as the HF data source. The corresponding governing equations for these two fidelities of flow field (convective form without energy equation) are shown below:

$$Euler\ flow: \begin{cases} \frac{D\rho}{Dt} = -\rho\nabla \cdot \boldsymbol{u} \\ \frac{D\boldsymbol{u}}{Dt} = -\frac{\nabla p}{\rho} + \boldsymbol{g} \end{cases} \tag{17}$$

$$Navier-Stokes\ flow: \begin{cases} \frac{D\rho}{Dt} = -\rho\nabla \cdot \boldsymbol{u} \\ \frac{D\boldsymbol{u}}{Dt} = -\frac{\nabla p}{\rho} + \frac{\mu\nabla^2 \boldsymbol{u}}{\rho} + \frac{\mu\nabla(\nabla \cdot \boldsymbol{u})}{3\rho} + \boldsymbol{g} \end{cases} \tag{18}$$

where $\rho$ is the air density, $t$ the time, $p$ the pressure, $\boldsymbol{u}$ the flow velocity vector, $\nabla$ the differential operator, $\mu$ the dynamic viscosity, $\boldsymbol{g}$ the body force, $\frac{D}{Dt}$ the material derivative. The 2D Euler flow was solved by the numerical code MISES (Multiple Blade Interacting Streamtube Euler Solver) developed by Drela and Youngren from MIT [51]. The mesh was generated by MISES/ISET and $y^+$ was set near 1 with around 3,180 cells generated in the LF simulation. It should be noted that the calculation of $y^+$ in this context is solely used for determining the spacing of the first layer of the mesh. In general, the solution of the Euler equations is not significantly affected by the grid density. The detailed 2D mesh inside the LS89 passage and the contour of Mach number are shown in Fig. 9.

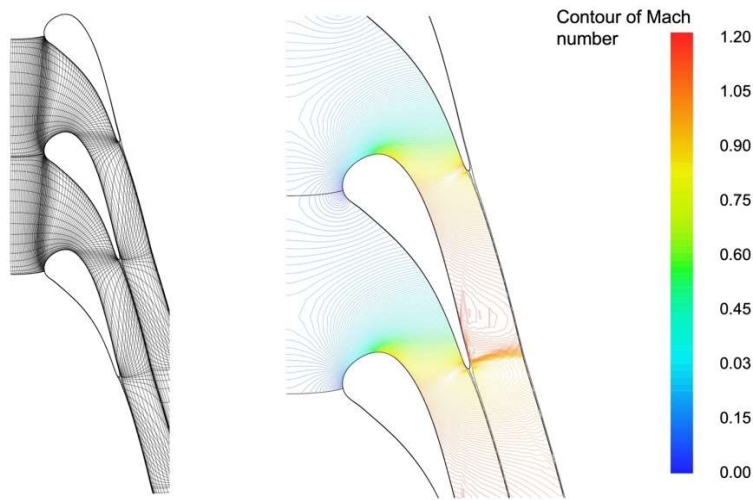

Fig. 9 Generated mesh and contour of Mach number inside turbine vane passages

The Euler flow result shows that there are two shock waves generated inside the vane passage, the front one due to the choking effect in throat area; the rear one near the trailing edge because of the flow acceleration on suction surface. However, the MISES generally specifies the trailing edge flow based on the Kutta condition instead of explicitly solving the flow physics [51]. As a result, the Euler flow results do not predict proper base pressure and losses in the trailing edge area [52-53]. Thus, the logic of the MF-DNN is to capture the basic flow pattern from LF Euler results and then to locally correct the trailing edge flow field using HF experimental measurement data.

Currently, the predictions primarily focus on the distributions of the isentropic Mach number by utilizing a combination of 2D Euler results and a limited number of experimental data points on the vane wall surfaces. It is crucial to acknowledge that this approach provides a partial prediction of the flowfield. To fully construct the complete flowfield using MF-DNNs, obtaining high-fidelity experimental data within the cascade passage is necessary. 160 LF points and 4 HF points were selected in total as the input training data for the MF-DNN. The near-wall flow fields on the pressure side and suction side of LS89 vane were modelled separately. The LF-DNN consisted of 5 hidden layers with 256, 128, 128, 64 and 32 neurons on each layer, respectively. The Correction DNN consisted of 4 hidden layers with 128, 64, 32 and 32 neurons on each layer, respectively. The comparison results are shown in Fig. 10. Instead of arranging pressure probes in the whole chordwise range of vane surface, the MF-DNN well-predicted the distributions of isentropic Mach number based on the 2D Euler flow field and 4 experimental measurement data points. The MF-DNN inherited the Euler flow pattern in the upstream zone and corrected the static pressure distributions near the shock wave occurring close to the trailing edge.

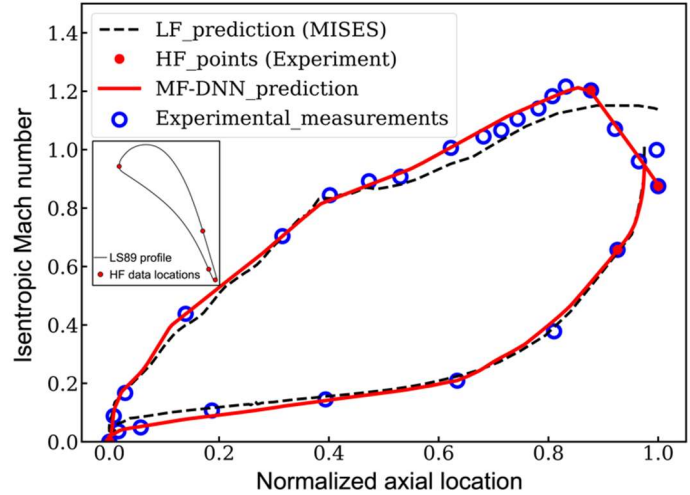

Fig. 10 MF-DNN prediction of near-wall isentropic Mach number on LS89 vane surface

## 4. Conclusions

A new architecture of multi-fidelity deep neural network (MF-DNN) was proposed in this paper to learn the training data with different fidelities. The performance of the built MF-DNN was then deployed in several benchmark tests of surrogate modelling and uncertainty quantification (UQ) for both the low-dimensional and high-dimensional functions. The main conclusions in this research work can be drawn as follows:

1) In the proposed MF-DNN, a single correction network was constructed here to autonomously learn either the linear or nonlinear correlation. The surrogate modelling capability of MF-DNN was tested by approximating the 1-, 32-, and 100-dimensional benchmark functions. The proposed MF-DNN exhibits comparable performance to the previous composite architecture in handling the nonlinear relationship between LF- and HF-data, while it outperforms the composite architecture when dealing with linear relationship.

2) The proposed MF-DNN demonstrated superior performance when compared to the traditional non-neural network models, particularly in handling high-dimensional problems. The performance of MF-DNN was evaluated in solving aleatory UQ problems, both in 1-, 32-and 100-dimensional settings, under input uncertainties following uniform and Gaussian distributions. MF-DNN demonstrated its efficiency in accurately predicting probability density distributions and statistical moments of quantities of interest (QoI).

3) MF-DNN was deployed to predict the distributions of near-wall isentropic Mach number on turbine vane LS89 by leveraging the 2D Euler results and a limited number of experimental measurement data points. The results demonstrated that the near-wall flow field was well-predicted by MF-DNN. In details, MF-DNN inherited the Euler flow pattern in the upstream zone and corrected the static pressure distributions near the shock wave using experimental data.

## Declaration of competing interest

The authors declare that they have no known competing financial interests or personal relationships that could have appeared to influence the work reported in this paper.

**Data availability**

Data will be made available on request.

**Acknowledgement**

The authors would like to sincerely thank the support of the grant from the European Union's Marie Skłodowska-Curie Actions Individual Fellowship (MSCA-IF-MENTOR-101029472).

**Nomenclature**

| | |
|---|---|
| DNN | deep neural network |
| GPU | graphics processing unit |
| HF | high-fidelity |
| KRG | Kriging model |
| LF | low-fidelity |
| MF | multi-fidelity |
| MF-DNN | multi-fidelity deep neural network |
| MISES | multiple blade interacting streamtube Euler solver |
| MSE | mean square error |
| NN | neural network |
| QoI | quantities of interest |
| RBF | radial basis function |
| UQ | uncertainty quantification |
| "ReLU" | rectified linear unit activation function |
| "Sigmoid" | logistic sigmoid function |
| "Tanh" | hyperbolic tangent function |
| $H$ | number of hidden layers |
| $M$ | number of LF training points |
| $N$ | normal (Gaussian) distribution |
| $P$ | number of HF training points |
| U | uniform distribution |
| $L_c$ | overall mean square error of the Correction DNN |
| $L_h$ | training loss of Correction DNN |
| $L_l$ | training loss of LF-DNN |
| $L_r$ | $L_2$ regularization loss of Correction DNN |
| $Q_{LF}$ | low-fidelity data set |
| $Q_{HF}$ | high-fidelity data set |
| $X_{LF}$ | low-fidelity variables |
| $X_{HF}$ | high-fidelity variables |
| $b_h$ | biases of $h$-th layer |
| $\hat{y}_{HF}$ | prediction values on HF data points |
| $y_{LF}$ | label values of LF data points |
| $\theta_{NN}$ | neural network parameters |
| $w_h$ | weights of $h$-th layer |
| $\mathcal{F}$ | correlation between LF- and HF data |
| $\lambda$ | control parameter of $L_2$ regularization loss |

| $\rho$ | multiplicative correction surrogate |
|---|---|
| $\delta$ | additive correction surrogate |

## Appendix A. Comparison of activation functions in neural network

The capability comparison of the different activation functions, i.e., the "ReLU", the "Sigmoid" and the "Tanh" in regression was conducted here. Theoretically, these three activation functions are non-linear based on the definition of non-linearity (i.e., the derivative of the dependent variable to the independent variable changes with the independent variable), and they have shown wide applications in approximating the non-linear functions. Their capability in approximating both the linear and nonlinear function was studied in this section. The simple linear/nonlinear unary functions are chosen as the governing equations as follows:

$$y = 2x, x \in [-1,1] \tag{A.1}$$

$$y = 2x^2, x \in [-1,1] \tag{A.2}$$

$$y = 2x^3, x \in [-1,1] \tag{A.3}$$

$$y = 2\sin(4x), x \in [-1,1] \tag{A.4}$$

Three neural networks (NNs) were established here, and each NN had a different activation function (i.e., "ReLU", "Sigmoid" and "Tanh", respectively). To exclude the influence of the other factors, the hyperparameters and initialization of the established NNs remained the same, and the only variable was the activation function within the training process. In detail, there were 21 training points in total that were generated uniformly distributed within the interval $x \in [-1, 1]$. The architecture of the NNs consisted of 2 hidden layers with 20 neurons on each layer. The training epoch number was 1000 and the batch size for each epoch was 7. Figure A.1 illustrates the comparison between the predictions obtained from NNs using different activation functions: "ReLU," "Tanh," and "Sigmoid." It is evident that both the "ReLU" and "Tanh" activations exhibit better agreement with the theoretical values when compared to the "Sigmoid" activation. To provide a quantitative assessment, Table A.1 presents the MSE values on the validation dataset. Across a wide range of linear and nonlinear functions, the "ReLU" activation function consistently demonstrates superior prediction accuracy compared to the other activation functions. These results reinforce the notion that the proposed Correction DNN with the "ReLU" activation function can effectively approximate both linear and nonlinear correlations across a broad spectrum of scenarios.

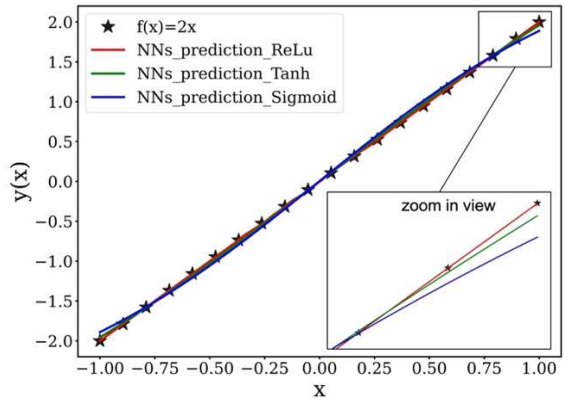

(a)

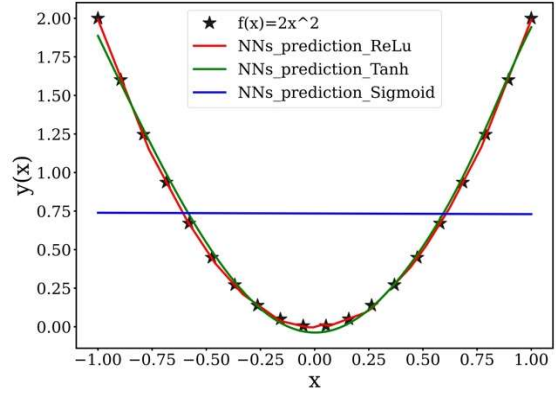

(b)

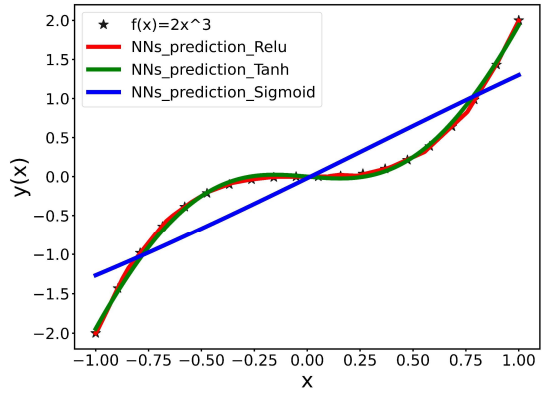

(c)

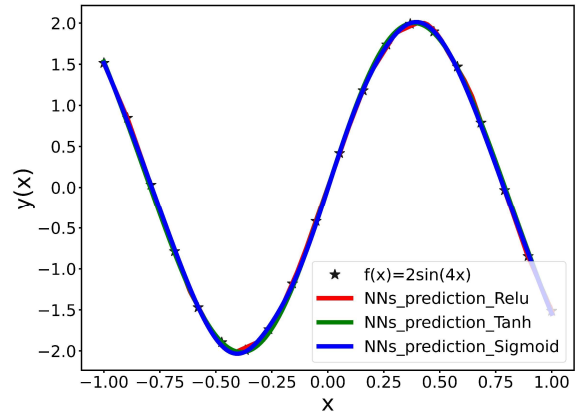

(d)

Fig. A.1 Comparison of various activation functions in approximating linear/nonlinear functions

Table A.1 Performance comparison of various activation functions

|  | ReLu | Tanh | Sigmoid |
| --- | --- | --- | --- |
| Test MSE for A.1 | 9.48E-06 | 4.50E-04 | 2.35E-03 |
| Test MSE for A.2 | 5.13E-05 | 1.25E-03 | 3.61E-01 |
| Test MSE for A.3 | 9.77e-05 | 1.38e-03 | 1.00e-02 |
| Test MSE for A.4 | 2.62e-04 | 1.12e-03 | 5.62e-04 |